\documentclass[prl,aps,amssym,nofootinbib,floatfix,reprint,notitlepage]{revtex4-1} 
\usepackage{wrapfig}
\usepackage{xcolor}
\usepackage{amsmath,amssymb,bbold,bm}
\usepackage{graphicx}
\usepackage[normalem]{ulem}
\usepackage{cancel}
\usepackage{comment}
\usepackage{hhline}

\newcommand{\sbraket}[1]{\langle #1 \rangle}
\newcommand{\hh}{\hphantom{.}}
\newcommand{\be}{\begin{equation}}
\newcommand{\ben}{\begin{equation*}}
\newcommand{\ee}{\end{equation}}
\newcommand{\een}{\end{equation*}}
\newcommand{\bs}{\begin{split}}
\newcommand{\es}{\end{split}}
\newcommand{\bmx}{\begin{array}}
\newcommand{\emx}{\end{array}}

\newcommand{\bea}{\begin{eqnarray}}
\newcommand{\bean}{\begin{eqnarray*}}
\newcommand{\eea}{\end{eqnarray}}
\newcommand{\eean}{\end{eqnarray*}}
\newcommand{\dg}{^{\dagger}}
\newcommand{\dn}{^{\vphantom{\dagger}}}

\newcommand{\lr}{\leftrightarrow}

\newcommand{\ua}{\uparrow}
\newcommand{\da}{\downarrow}

\newcommand{\bb}[1]{\mathbb{#1}}
\newcommand{\qqquad}{\qquad\qquad\qquad}
\newcommand{\so}{\qquad\rightarrow\qquad}

\newcommand{\orr}{\qquad\text{or}\qquad}
\newcommand{\andd}{\qquad\text{and}\qquad}

\newcommand{\eps}{\epsilon}

\newcommand{\sgn}[1]{{\rm sign}{#1}}
\newcommand{\pref}[1]{(\ref{#1})}

\newcommand{\ket}[1]{\left\vert #1\right\rangle}

\newcommand{\matl}[1]{\bmx{ll}#1\emx}

\newcommand{\bw}[1]{\begin{widetext}}
\newcommand{\ew}[1]{\end{widetext}}

\newcommand{\gray}[1]{}

\newcommand{\nothing}[1]{}

\begin{document}

\title{Topological ground state degeneracy of the two-channel Kondo lattice}
\author{Aleksandar Ljepoja}
\author{Yashar Komijani\,$^{*}$}
 \affiliation{ Department of Physics, University of Cincinnati, Cincinnati, Ohio, 45221, USA}
\date{\today}
\begin{abstract}
There are indications from the large-N analysis that multi-channel Kondo lattices have topological order. We use the coupled-wire construction to study the channel paramagnetic regime of a two-channel Kondo lattice model of spin-1/2 SU(2) spins. Using abelian bosonization we show that in presence of particle-hole symmetry, each wire is described by a [SO(5)$\times$Ising]/Z$_2\times$ SU(2) symmetric theory. When the wires are coupled together and the time-reversal symmetry is broken, the system exhibits topological order with fractional edge states and anyonic excitations. By an explicit construction of the Heisenberg algebra acting on the ground state manifold, we demonstrate that in presence of particle-hole symmetry, the ground state on a torus is eight-fold degenerate. This is also discussed using a heuristic approach which is applicable to other topologically ordered states.
\end{abstract}
\maketitle

\emph{Introduction} --
The Kondo effect, originally formulated to describe the screening of localized magnetic impurities by conduction electrons, has long served as a cornerstone in condensed matter physics \cite{Hewson,Si2014,Coleman2015}. The two-channel Kondo (2CK) effect \cite{Nozieres80,Andrei84,Affleck92,Emery1993,Affleck1993}, which defies the conventional Fermi liquid paradigm, has attracted significant attention due to its non-Fermi liquid fixed point (FP). While much progress has been made in understanding 2CK effects at the impurity level, including their potential application in topological quantum computation (TQC) \cite{Lopes20,Komijaniqubit,Ren2024}, the collective behavior of 2CK lattice (2CKL) systems \cite{Jarrell96,Jarrell1997,Cox1996,Schauerte05,Hoshino11,Kuramoto,vanDyke19,Zhang18,Inui20,Flint2024} remain less explored. Recent work using large-N expansions \cite{Wugalter2020,Rebecca21,Ge2022,Ge2024} suggests that multi-channel Kondo lattices can exhibit topological order \cite{Wen1990}, potentially hosting non-Abelian anyonic excitations and topologically protected edge states. This and the prospect of realizing them in quantum materials \cite{Onimaru2019,Patri2020} makes them potentially relevant for TQC \cite{Ge2024,Rebecca21}.

In the \emph{2CK impurity model} two spin-full electron channels are Kondo coupled to a magnetic impurity and compete to screen it. The two spin-full channels correspond to a collection of eight chiral Majorana modes, which in presence of particle-hole symmetry (PHS) transform under the SO(8) symmetry. When the Kondo coupling strengths are equal this leads to an intermediate coupling FP with a residual entropy of $\frac{1}{2}\log 2$. The abelian bosonization route \cite{Emery1993} shows that the latter arises from a decoupled Majorana zero mode (MZM) at the IR FP. Furthermore, coupling to the impurity, changes the boundary condition of three of these chiral Majorana modes 
\cite{Maldacena1997,Ye1997}, reducing the symmetry to SO(5)\,$\times$\,SO(3). The SO(5)\,$\sim$\,SP(4) symmetry arising from the decoupled U(1) charge and SU(2) channel sectors \cite{Affleck92}, characterize all relevant perturbations to the IR FP. Remarkably, the decoupled MZM, hereafter referred to as  $\gamma'$, has been elusive in all but thermodynamic calculations. 

\begin{figure}[tp]
\includegraphics[width=\linewidth]{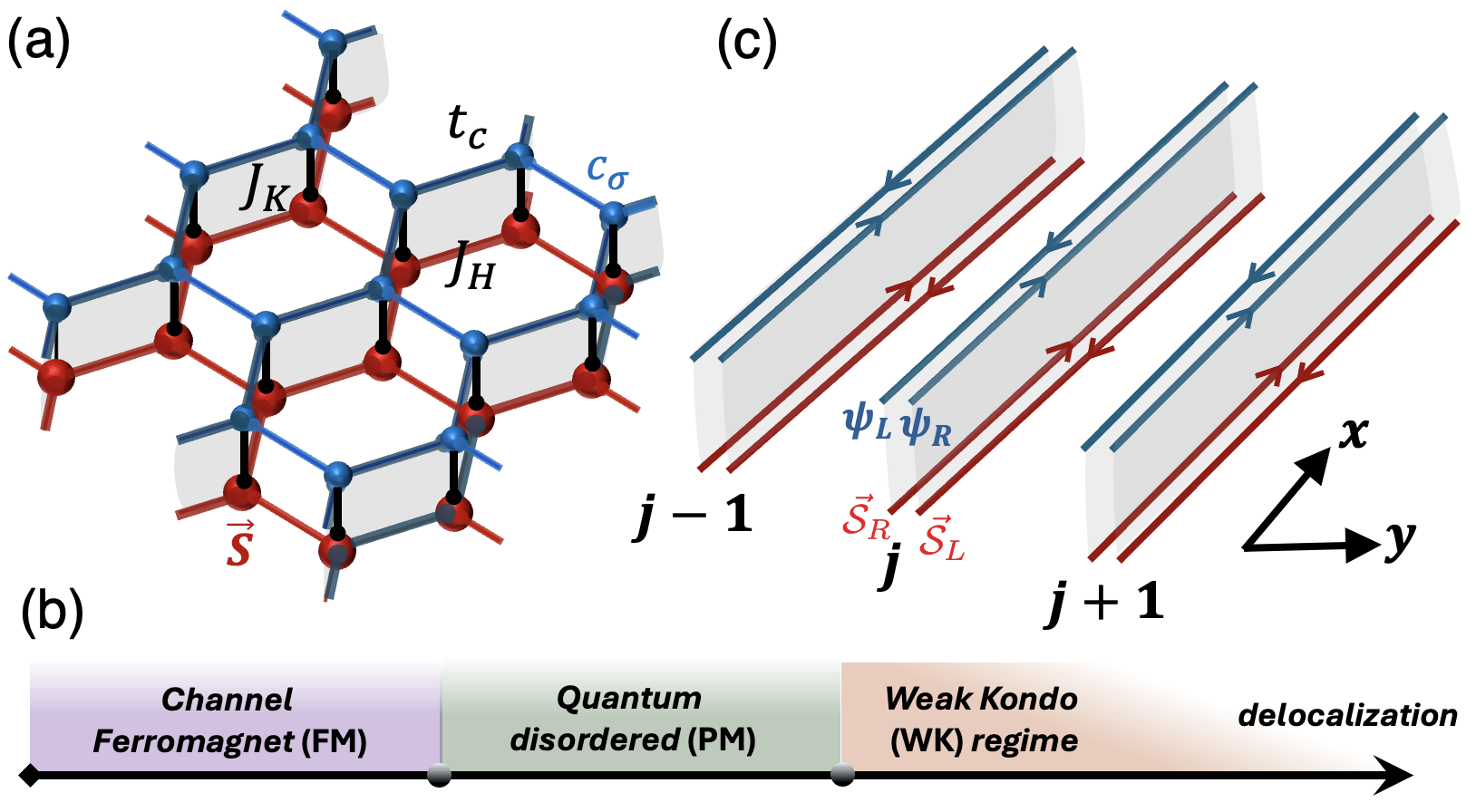}
\caption{(a) The two-channel Kondo lattice (2CKL) on a honeycomb lattice. (b) The phase diagram in absence of TRS from dynamical large-N \cite{Ge2024}. (c) The coupled-wire construction (CWC) of the model in (a). Each bundle $j$ is composed of a chiral $j$ and an anti-chiral $\bar j$ wires. We assume periodic boundary condition in both directions, realizing a torus geometry.}\label{fig:Fig1}
\end{figure}

Recently, the 2CKL model has been studied in both 1D \cite{Ge2022} and 2D \cite{Ge2024} in the limit of large spin group and large channel number. These studies indicate the emergence of Lorentz and scale invariance as well as velocity-locking from the lattice model, so that the IR FP is governed by a channel-symmetric conformal FP. In particular in 2D, breaking time-reversal symmetry (TRS) leads to three phases [Fig.\,\ref{fig:Fig1}(b)] including channel ferromagnetic (FM) regime where the symmetry is spontaneously broken, the channel paramagnetic (PM) regime, where defects restore the symmetry and a weak Kondo-coupling (WKC) regime. In the latter phase, the TRS breaking is induced on the spinons \cite{Ge2022,Ge2024,Shao2024}, creating a chiral spin liquid, whose edge states run in the opposite direction to electrons. The edge states are Kondo coupled leading to the same coset theory \pref{eqOA} on the boundary. It is unclear how much of these results extend to the SU(2) 2CKL. The goal of this paper is to answer this question.

In this paper, we focus on the 2CKL with spin-1/2 SU(2) local moments, studied using the coupled-wire construction (CWC) approach \cite{Kane02,Teo14}, which has been successfully applied to various strongly correlated systems \cite{Kane17,Sirota19,Fuji19,Imamura19,Meng15,Patel16,Pereira18,Iadecola2016,Iadecola2019,Bomantara19,Mross17,Li20,BuenoXavier2023}. In this quasi-1D framework, we will show that many of the elements of the 2CK impurity re-appear in the CWC approach to the 2CKL. In particular $\gamma'$ will appear as a chiral mode, whose elusiveness is rooted in the gauge charge it carries. Our work highlights the topological nature of the 2CKL and the existence of fractionalized edge states. We demonstrate that the topological ground state degeneracy (GSD) on a torus geometry \cite{Moore1991,Read1999,Read2000,Kitaev2003,Kitaev2006} can be calculated from the explicit construction of a Heisenberg algebra \cite{Oshikawa2007,Iadecola2019} acting on the ground state manifold. This provides a heuristic framework for understanding the interplay between symmetry breaking and topological order in 2CK Kondo lattice systems, with potential implications for other topologically ordered states.

The \emph{1D 2CK lattice model} has been studied using both non-abelian  \cite{Andrei2000} and abelian 
\cite{Emery1993,Azaria1998,Azaria2000,Azaria2000b} bosonization techniques. The chiral spin currents of electrons and magnetic moments, represented by $\vec{\cal J}$ and $\vec{\cal S}$ and obeying SU(2)$_{2}$ and SU(2)$_{1}$ Kac-Moody (KM) algebra, are coupled via Kondo interaction $J_K(\vec{\cal J}_R\cdot\vec{\cal S}_L+\vec{\cal J}_L\cdot\vec{\cal S}_R)$. The KM level mismatch leads to a \emph{chirally-stabilized} \cite{AJZ,Andrei2000} FP, whose spin-sector at the IR, is described by
\be
[{\rm SU(2)}_{1}\times\overline{\rm Ising}]\otimes[\overline{\rm SU(2)}_{1}\times{{\rm Ising}}],\label{eqOA}
\ee
in addition to the decoupled charge and channel sectors. The Ising conformal field theory (CFT), appearing as a coset model in \cite{Andrei2000}, contains the 1D version of the $\gamma'$.

\begin{figure}[t]
\includegraphics[width=\linewidth]{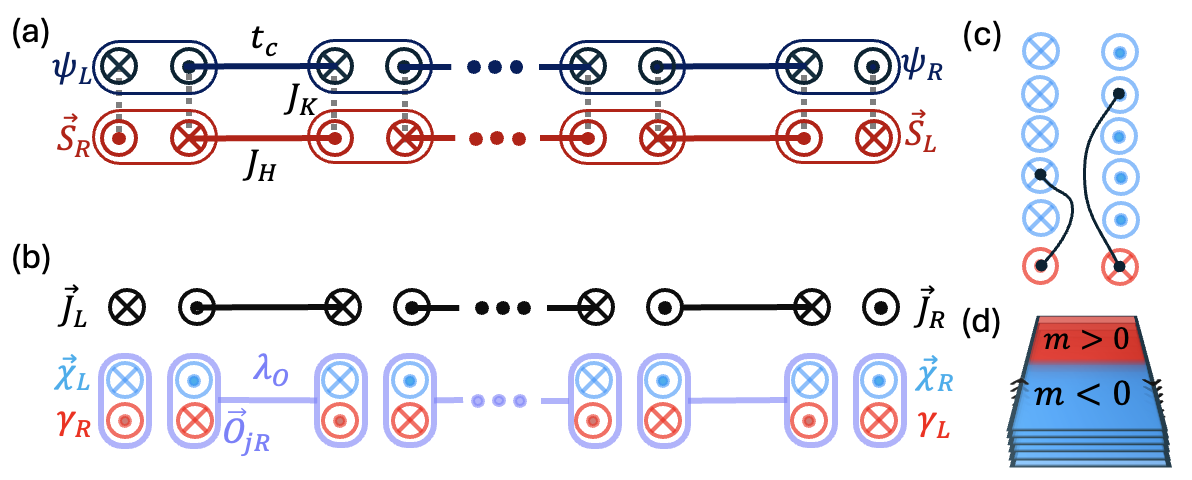}
\caption{The CWC setup (a) The weak Kondo coupling (WKC) limit (b) The strong Kondo coupling limit is a quantum paramagnet (PM) (c) SO(5) symmetry broken phase (FM) (d) topological defects are in-phase domain walls in the mass gap of each patch.}
\label{fig:Fig2}
\end{figure}

\emph{The Model} -- Since the 1D 2CKL is understood \cite{Emery1993,Andrei2000,Azaria2000}, we construct an anisotropic limit of the 2D 2CKL using CWC. The Hamiltonian can be written as
\be
H=\sum_j\int{dx}{\cal H}_j, \quad {\cal H}_j={\cal H}_j^{\rm (sp)}+{\cal H}_j^{\rm (el)}+{\cal H}_j^{\rm(hop)}+{\cal H}_j^{\rm(K)},\nonumber
\vspace{-.25cm}
\ee
where each {bundle} $(j,\bar j)$ is composed of a Heisenberg spin-chain, represented by SU(2)$_1$ KM currents $\vec {\cal S}_{j,R/L}$ and chiral electrons $\psi_{j,R/L,a\alpha}$, described by
\bea
{\cal H}^{\rm (sp)}_j&=&\frac{2\pi v}{6}(\vec {\cal S}_{j,R}^2+\vec {\cal S}_{j,L}^2)\\
{\cal H}^{\rm (el)}_j&=&-iv\sum_{\mu\alpha}(\psi\dg_{j,R\mu\alpha}\partial_x\psi\dn_{j,R\mu\alpha}-\psi\dg_{j,L\mu\alpha}\partial_x\psi\dn_{j,La\alpha}).\nonumber
\vspace{-.25cm}
\eea
Here, $\alpha=\ua,\da$ and $\mu=1,2$ are channel indices, respectively.
The WKC limit has a natural CWC description (Fig.\,\ref{fig:Fig1}a). A brick wall version of Haldane model \cite{Jotzu2014} can be broken down to weakly coupled wires  (Fig.\,\ref{fig:Fig1}c), with inter-wire coupling $\sum_{j,n}(t_n\psi\dg_{R,j}\psi\dn_{L,j+n}+h.c.)$. We assume $t_n\neq 0$ only for $n=\pm1$. For $t_1=t_{-1}$ this gives two Dirac points, but due to broken TRS, one of the $t_{\pm1}$-s dominate, gapping out the bulk except for a left/right-mover at the two ends, similar to the Su-Schrieffer-Heeger model \cite{Su1979}. Likewise, in the magnetic layer, each spin-chain is described by a SU(2)$_1$ WZW model and inter-wire AFM couplings lead to $\sum_{j,n}J^H_{n}\vec S_{j,R}\cdot\vec S_{j+n,L}$. Due to Kondo-induced TRS breaking \cite{Ge2024,Shao2024}, again one of $J^H_{\pm 1}$-s dominate, gapping the bulk \cite{Meng15,Patel16,Iadecola2019}. Eventually, a chiral (anti-chiral) spinon (electron) is left at each edge, Kondo-coupled as in the WKC discussion above [Fig.\,\ref{fig:Fig2}(a)]. Therefore, we will pursue an extreme TRS-broken version of the model, where one chirality is chosen
\ben
\hspace{-.1cm}{\cal H}_{j}^{({\rm hop})}\hspace{-.15cm}= \sum_{\mu\alpha} t(\psi\dg_{j,R\mu\alpha} \psi\dn_{j+1,L\mu\alpha}+h.c.) + J_{H} \vec {\cal S}_{,jL}\cdot\vec {\cal S}_{j+1,R}.
\een
The chiral fermions $\psi_{j,L/R,\mu\sigma}$ have a Sugawara decomposition \cite{Francesco1997} into U(1)$\times$ SU$^{\rm sp}$(2)$_2\times$SU$^{\rm ch}$(2)$_2$ fields. Defining the SU(2)$_2$ currents $\vec{\cal J}=\sum_{\mu,\alpha\beta}\psi\dg_{\mu\alpha}\vec\sigma_{\alpha\beta}\psi\dn_{\mu\beta}$, the Kondo coupling decouples \cite{Andrei2000} into an RL and an LR sectors [Fig.\,\ref{fig:Fig1}(c)]:
\bean\label{eq:hamiltonian}
  {\cal H}_{j}^{(K)} \hspace{-.1cm}=
[\lambda^{z}_{K} {\cal J}^{z}_{L} {\cal S}^{z}_{R} + \frac{\lambda^{\perp}_{K}}{2} \big( {\cal J}_{L}^{+} {\cal S}_{R}^{-} + {\cal S}_{R}^{+} {\cal J}_{L}^{-} \big )]+(R\lr L).\label{eqHK}
\eean
This anisotropic model enables using abelian bosonization to solve the intra-wire problem exactly \cite{Azaria2000} at the so-called Toulouse point for a particular $\lambda^z_K/\lambda^\perp_K$, but the full SU(2) symmetry is restored at the IR FP \cite{Lotem2024}.  

\emph{Intra-wire IR fixed point} -- We review this part \cite{Azaria2000} to set the notation and generalize the discussion to the SO(5) symmetric version. Within abelian bosonization, $\psi_{L,\mu \alpha} \sim e^{i\phi_{\mu \alpha}}$ and $\psi_{R,\mu \alpha} \sim e^{i\bar\phi_{\mu \alpha}}$, after a standard rotation of bosons \cite{vonDelft1998} to charge $\phi_c$, spin $\phi_s$, flavor $\phi_f$ and spin-flavor $\phi_{sf}$ bosons, the currents can be expressed as:
\be
{\cal S}_R^{\pm} \sim e^{\mp i \sqrt{2} \bar{\phi}_{m}},\qquad  {\cal J}_L^{\pm} \sim  e^{\mp i \phi_{s}}\cos(\phi_{sf})
\ee
where the Klein factors have been ignored since they do not play an important role here \cite{SM}. This allows us to rewrite the Kondo interaction in a bundle in terms of bosonic degrees of freedom. We use $\psi=e^{i\phi}=\chi'+i\chi''$ \cite{Ftnote2}. Focusing on LR sector (RL is similar) 
\be
\big({\cal J}_{L}^{+} S_{R}^{-} +{\cal J}_{L}^{-}S_{R}^{+} \big ) \to \big( e^{-i\sqrt{2}\bar{\phi}_{m}} e^{-i \phi_{s}} + h.c. \big ) \cos(\phi_{sf}). \nonumber
\ee
Using a non-trivial canonical transformations $\bar{\phi}_{\gamma} \equiv \sqrt{2} \bar{\phi}_{m} + \phi_{s}$ and $\phi_{J} \equiv \sqrt{2} \phi_{s} + \bar{\phi}_{m}$ simplifies this interaction, and also removes the longitudinal ${\cal J}^zS^z$ interaction \cite{Azaria2000,SM} at the Toulouse point. Reconstructing $J^+=e^{-i\sqrt{2}\phi_J}$
and re-fermionizing $e^{i\phi_\gamma}=\gamma'+i\gamma''$ in terms of Majoroana modes, leads to:
\begin{equation}
\label{eq:gapped}
\lambda_K^\perp i\sin(\bar{\phi}_{\gamma}) \cos(\phi_{sf}) = \lambda_{\perp} i\bar{\gamma}'' \chi^{\prime}_{sf}
\end{equation}
This interaction gaps out the two chiral Majoranas \cite{SM},  $\langle i\bar\gamma'' \chi'_{sf} \rangle =1/ \lambda_K^\perp$, reducing the central charge of each wire from $c_{\rm UV}=4+\bar 1$ with U(2k) $\otimes$ SU$_{1}$(2) KM algebra to $c_{\rm IR}=5/2+1+\overline{1/2}$ with SO(5)$_1\otimes$ SU$_{1}$(2) KM algebra plus an additional chiral Majorana model $\bar\gamma'$. The decoupled Majoranas transforming under SO(5) are $\chi^a=\{\chi^{\prime}_{c}$, $\chi^{\prime \prime}_{c}$, $\chi^{\prime}_{f}$, $\chi^{\prime \prime}_{f}$, $\chi^{\prime \prime}_{sf}\}$. Additionally at the IR we also have an SU(2)$_{1}$ current $J^{\pm} \sim e^{\mp i \sqrt{2}\phi_{J}}$ that is free and decoupled from the rest. In other words, at the IR $\vec{\cal J}_L$ and $\vec {\cal S}_R$ combine to produce a SU(2)$_1$ current $\vec J$, and the Ising fermion $\bar\gamma'$, as indicated by \pref{eqOA}. However, the IR FP turns out to be slightly more complex, as we discuss next.

\emph{Gauge symmetry} -- There is a subtlety involving the gauge invariance of the remaining Majorana modes. In the original basis with electrons $\psi_{\mu\alpha}\sim e^{i\phi_{\mu\alpha}}$ and the spin current ${\cal S}_{R}^{\pm} \sim e^{\mp i \sqrt{2} \bar{\phi}_{m}}$, the bosons have compactification radius of $\phi_{a\alpha}\sim\phi_{a\alpha}+2 \pi n_{a\alpha}$ and $\bar{\phi}_{m} \to \bar{\phi}_{m}+\sqrt{2}\pi \bar n_m$ for arbitrary integers $n_{a\alpha}$ and $\bar n_m$. At the IR, bosonic degrees of freedom transform as:
\bea
\phi_{c} &\to& \phi_{c} + \pi (n_{1\ua} + n_{1\da}+n_{2\ua} + n_{2\da})\nonumber\\
\phi_{f} &\to& \phi_{f} + \pi (n_{1\ua} + n_{1\da}-n_{2\ua} - n_{2\da})\nonumber\\
\phi_{sf} &\to& \phi_{sf} + \pi 
(n_{1\ua} - n_{1\da}-n_{2\ua} + n_{2\da})
\nonumber\\
\bar\phi_{\gamma} &\to& \bar\phi_{\gamma} + \pi(2\bar n_m - n_{1\ua}+n_{1\da}-n_{2\ua}-n_{2\da})\nonumber\\
\phi_{J} &\to& \phi_{J} + \pi\sqrt{2}(n_{1\ua}-n_{1\da}-n_{2\ua}+n_{2\da}-\bar n_m).
\eea
Changing any of $n_{a \alpha}$ by one unit changes the sign of all remaining Majorana fermions, indicating an emergent $Z_2$ gauge symmetry. This has been discussed in the context of the impurity problem before \cite{BenTov2015}, and physically, it stems from gluing conditions that require collective excitations to change simultaneously in pairs. This leaves only the following gauge-invariant operators in each wire:
\be
{\cal O}^{a}_{j} =i \bar{\gamma}'_{j} \chi^{a}_{j}, \qquad 
{\cal I}_j^{ab} =i \chi^{a}_{j} \chi^{b}_{j}, \qquad J^+=e^{-i\sqrt{2}\phi_{J}}.\label{eqOIJ}
\ee
The ${\cal I}^{ab}$ is the chiral current of SO(5)$_1$, and $\vec J$ is the emergent SU(2)$_1$ current. The ${\cal O}$-s are relevant operators that can destabilize the IR fixed point, and beside channel symmetry breaking include composite pairing superconductivity. This has been studied in connection to the 2CK model \cite{Azaria2000b,Toth2008,Flint2011,Flint2014,Hoshino2014,Hoshino2014b,Kuramoto2016}  and can be interpreted, using $i\partial_t\psi\dg_{R\mu\ua}\psi\dg_{L\mu\da}=[\psi\dg_{R\mu\ua}\psi\dg_{L\mu\da},H]$, as odd-frequency pairing \cite{Berezinskii74,Abrahams95,Bergeret05,Linder19}. These operators, along with their microscopic expressions and dimensions are listed in table~\ref{table:tab1}. Their scaling dimension flows from UV to $\Delta=1$ at the IR. 

\emph{Inter-wire coupling} -- In terms of the IR fields \pref{eqOIJ}, the leading-relevant inter-wire Hamiltonian can be written as
\bea
{\cal H}_j&\to&\frac{2 \pi v}{6}(J^{2}_{j,R}+J^{2}_{j+1,L}) + \lambda_{J}\sum_{j}\vec{\bar J}_{j}\cdot \vec J_{j+1}\label{eqHOIJ} \\
&&\hspace{-1cm}+iv\sum_a[(\bar\chi^a_j\partial_x\bar\chi^a_j-\chi^a_{j+1}\partial_x\chi^a_{j+1})+(\gamma_j\partial_x\gamma_j-\bar\gamma_{j+1}\partial_x\bar\gamma_{j+1})]\nonumber\\
&+& \lambda_{\cal O} \sum_{a}\sum_{j} \bar{\cal O}^{a}_{j}{\cal O}^{a}_{j+1} + \lambda_{I} \sum_{j} \sum_{a,b} \bar{\cal I}^{ab}_{j} {\cal I}^{ab}_{j+1}.\nonumber
\eea
 This effective model is valid in the strong Kondo-coupling limit and is schematically shown in Fig.\,\ref{fig:Fig2}(b). It can be derived using a perturbation theory in $t/\lambda_K$ and $J_H/\lambda_K$, resulting in { $\lambda_{I} \sim {t^2}/{\lambda_{K}}$ and $\lambda_{O} \sim {t^2 J_{H} }/{\lambda^{2}_{K}}$}. Note that the fields are paired into new bundles ($\bar j,j+1)$, hereafter referred to as \emph{superconducting patches}.

\begin{table}[tp]
\begin{tabular}{c|c|c}
$\vec {\cal O}$ & operator & $\Delta$\\
\hline
$i\bar\gamma'\chi_f' $ & $(\psi\dg_{L}\vec\sigma\tau^x\psi\dn_{L})\cdot\vec {\cal S}\dn_R$ & 1\\
$i\bar\gamma'\chi_f'' $ & $(\psi\dg_{L}\vec\sigma\tau^y\psi\dn_{L})\cdot\vec {\cal S}\dn_R$ & 1\\
$i\bar\gamma'\chi_{sf}'' $ & $(\psi\dg_{L}\vec\sigma\tau^z\psi\dn_{L})\cdot\vec {\cal S}\dn_R$ & 1\\
$i\bar\gamma'\chi_c' $ & $(\psi\dg_{L1}\vec\sigma\psi\dg_{L2}+h.c.)\cdot\vec {\cal S}\dn_R$ & 1\\
$i\bar\gamma'\chi''_c $ & $i(\psi\dg_{L1}\vec\sigma\psi\dg_{L2}-h.c.)\cdot {\cal S}\dn_R$ & 1
\vspace{.35cm}\\
$\vec J$ & operator & $\Delta$\\
\hline
$J^+$ & $\psi\dg_{L1\da}\psi\dn_{L1\ua}\psi\dg_{L2\da}\psi\dn_{L2\ua}{\cal S}^-_R$ & 1\\
$J^-$ & $\psi\dg_{L1\ua}\psi\dn_{L1\da}\psi\dg_{L2\ua}\psi\dn_{L2\da}{\cal S}^+_R$ & 1\\
$J^z$ & $\frac{1}{2}\psi\dg_{L}\sigma^z\psi\dn_{L}+{\cal S}_R^z$ & 1\\
\end{tabular}
\hspace{-.1cm}
\begin{tabular}{c|c|c}
$\vec{\cal I}$ & operator & $\Delta$\\
\hline
$i\chi'_{f}\chi''_{sf}$ & $(\Psi\dg_{L}\tau^x \eta^z\Psi\dn_{L})$ & 1\\
$i\chi''_f\chi''_{sf}$ & $(\Psi\dg_{L}\tau^y \eta^0\Psi\dn_{L})$ & 1\\
$i\chi'_f\chi''_f$ &$(\Psi\dg_{L}\tau^z \eta^z\Psi\dn_{L})$ & 1 \\
$i\chi'_c\chi''_c$ & $(\Psi\dg_{L}\tau^0 \eta^z\Psi\dn_{L})$ & 1\\
$i\chi'_c\chi'_f$ &$(\Psi\dg_{L}\tau^z \eta^y\Psi\dn_{L})$ & 1 \\
$i\chi'_{c}\chi''_f$ & $(\Psi\dg_{L}\tau^0 \eta^x\Psi\dn_{L})$ & 1\\
$i\chi'_c\chi''_{sf}$ & $(\Psi\dg_{L}\tau^x \eta^x\Psi\dn_{L})$ & 1\\
{$i\chi''_c\chi'_f$} & $(\Psi\dg_{L}\tau^z \eta^x\Psi\dn_{L})$ & 1 \\
$i\chi''_{c}\chi''_f$ & $(\Psi\dg_{L}\tau^0 \eta^y\Psi\dn_{L})$ & 1\\
$i\chi''_{c}\chi''_{sf}$ & $(\Psi\dg_{L}\tau^x \eta^y\Psi\dn_{L})$ & 1\\
\end{tabular}
\caption{\small List of gauge-invariant operators at the IR FP of individual wires. The $\vec{\cal I}$ and $\vec J$ are currents of the SO(5)$\otimes$ SU(2) FP. The quintuplet $\vec{\cal O}$ operators are relevant perturbations that break the SO(5) symmetry. The operators have dimensions $\Delta=2$ or 3 at the UV but become $\Delta_{IR}=1$ at IR FP. The isospin subspace is defined as $\Psi_{\eta=1}\equiv\psi$ and $\Psi_{\eta=2}\equiv i\sigma^y\psi^*$ is its particle-hole conjugate.\,$\tau^a$ and $\eta^a$ are Pauli matrices in channel and isospin space, respectively. Remarkably, despite the non-local (bosonic) transformation, bilinear of Majoranas remain bilinear in the new Majorana basis.}\label{table:tab1}
\end{table}

\begin{figure}[t!]
\includegraphics[width=\linewidth]{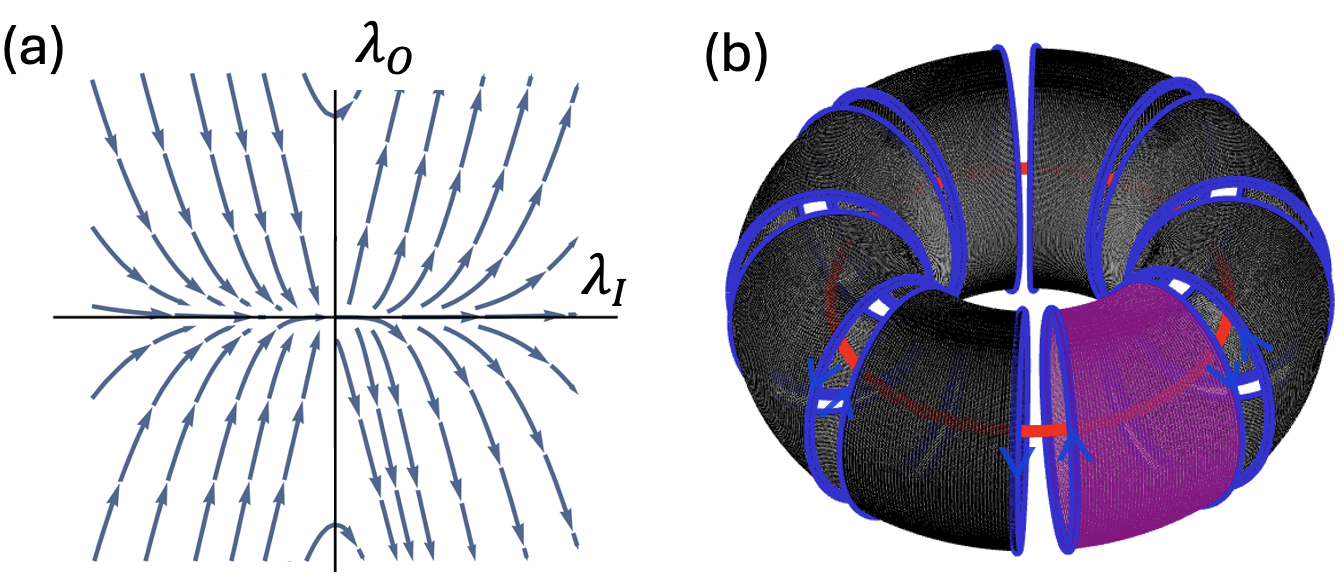}
\caption{(a) The RG flow of the coupling constants $\lambda_{O}$ and $\lambda_{I}$. For $\lambda_I>0$, both are relevant and the symmetry is restored.  (b) The black/purple shades denote trivial/topological superconducting patches between neighboring wires. The $\ket{++}$ state, containing an odd number of topological patches and a threading $\pi$-flux (red curve) does not belong to the ground state manifold for $N$ odd due to the opposite fermionic parity. }\label{fig:Fig3}
\end{figure}

\emph{Renormalization Group (RG) flow} -- 
To analyze the IR fate of the effective model \pref{eqHOIJ}, we study how  the coupling constants ($\lambda_{I}$, $\lambda_{O}$) evolve at long distances using an RG study. This can be done in the so-called Cardy scheme along with operator-product expansions of ${\cal O}(z)$ and ${\cal I}(z)$  operators, extracted from those of Majorana fermions. This enables us to obtain the RG flow \cite{SM} as:
\be
\frac{d \lambda_{I}}{dl} =  N\lambda^{2}_{I}+\lambda_O^2, \qquad 
\frac{d \lambda_{O}}{dl} = N \lambda_{O} \lambda_{I}
\ee
By studying the flow, shown in Fig.\,\ref{fig:Fig3}(a) we can see that for our choice of the signs of the coupling constants both flow towards strong coupling, reducing the anisotropy. Assuming $\lambda_O<0<\lambda_I$, this leads to $\lambda_O/\lambda_I=-1$.  

\emph{SO(N) generalization -- } At the RG FP, the Hamiltonian can be expressed as as ${\cal H}_j^{int}=-\lambda(i\Sigma_a\bar\chi^a_j\chi^a_{j+1}-i\gamma_j\bar\gamma_{j+1})^2$ and the symmetry group is enlarged.  By identifying $\bar\chi^6_j\equiv  \bar\gamma_{j+1}$ and $\chi^6_j\equiv  \gamma_{j-1}$, the Majorana modes can be combined into  $(\bar\chi^a_j,\bar\gamma_{j+1})$ and $(\chi^a_j,\gamma_{j-1})$, mapping the problem to the $N=6$ version of the model
\be
{\cal H}_j=iv\sum_a(\chi^a_{j+1}\partial_x\chi^a_{j+1}-\bar\chi^a_j\partial_x\bar\chi^a_j)-\lambda\big(\sum_ai\bar\chi^a_j\chi^a_{j+1}\big)^2.
\label{eqSON}
\ee
This model is SO(N) symmetric and invariant under $\chi\to-\chi$. The chiral interaction is exactly marginal for $N=2$, and marginally relevant for $N>2$. In the latter case, the spectrum is gapped by forming masses $m(x)\equiv\sum_a\sbraket{i\bar\chi^a_j(x)\chi^a_{j+1}(x)}\neq 0$. This can also be established by the mean-field theory in the $N\to\infty$ limit. Next, we study topological  defects and GSD of Eq.\,\pref{eqSON}.

\emph{Topological defects} are domain walls [Fig.\,\ref{fig:Fig2}(d)] in the emergent masses across which $\sbraket{i\bar\chi^a_j(x)\chi^a_{j+1}(x)}$ changes sign. According to Eq.\,\pref{eqSON}, the sign change has to be in-phase, happening simultaneously in all $N$ fields. Each defect harbors $N$ localized MZMs $\eta^a$ \cite{Tewari2007}. For even (odd) $N$, defects are bosonic (fermionic), respectively. 

\emph{Phase diagram -- } The CWC provides a description of both WKC and PM regimes. The SO(5) symmetry of the model with PHS can break spontaneously, resulting in $\sbraket{{\cal O}^a}\neq 0$. In the current quasi-1D scheme, however, the SO(5) symmetry is protected by the Mermin-Wagner theorem. In other words, the patches are decoupled and there is no stiffness in the entire system. This can be fixed by adding a term ${\cal H}_j\to{\cal H}_j+ \lambda'\bar{\cal O}^a_j{\cal O}^a_j$ that acts as a Weiss field between decoupled patches. This does not commute with the effective Hamiltonian \pref{eqHOIJ}, creating defects in nearby patches. These six $\eta^a$  MZMs create only four gauge-invariant states \cite{Ftnote1}, corresponding to a SP(4) spinorial boson. As $\lambda'/\lambda$ increases, a topological phase transition occurs from the topological $\sbraket{\bar{\cal O}^a_j{\cal O}^a_{j+1}}\neq 0$ to the trivial $\sbraket{\bar{\cal O}^a_j{\cal O}^a_{j}}\neq 0$ phase.\,At the vicinity of this transition,\,the proliferation of bosonic defects destroys the PM regime and their condensation leads to the FM phase.

\emph{Ground state degeneracy -- } In this section we analyze the GSD of the quantum PM on a torus geometry, using heuristic arguments, which are corroborated by an operators formalism in the next section. The decoupled SU(2)$_1$ has a GSD of 2, which we review in \cite{SM}. Hereafter, we focus on the remaining SO(6) theory. The sign of the mass gap determines the topological or trivial state of each $(\bar j,j+1)$ patch. However, due to $\chi^a\to-\chi^a$ gauge symmetry, this sign can be reversed at any time, without any energy cost. The only constraint is that due to the chiral anomaly, since $\chi_j$ and $\bar\chi_j$ have the same local origin, $\chi^a_j\to-\chi^a_j$ has to be accompanied by $\bar\chi^a_j\to-\bar\chi^a_j$ gauge transformation. Therefore, only the parity of the number of topological patches is gauge-invariant. Mapping each patch-flavor to an Ising chain, these correspond to ordered/disordered phases and therefore, we refer to this symmetry as the Kramers-Wannier (KW) \cite{Boyanovsky1989,Gogolin2004,Mussardo2009} topological symmetry of the patch in the following.

Additionally, each chiral Majorana fermion can have either periodic or anti-periodic boundary condition. This can be induced by the Aharonov-Bohm effect of a threading $\pi$-flux through the narrow hole of the torus. This has to be equal between both $\bar\chi_j$ and $\chi_{j+1}$, for the translational invariance along the patch to be maintained. This is the second topological degeneracy, resulting is the GSD of 4. However, the two are coupled. Passing a $\pi$-flux changes the parity of occupied states in topological patches. Therefore, in presence of an odd number of topological patches,  changing the boundary condition results in change of the occupation by $N$ electrons. Since the underlying theory is superconducting, only the total parity of charge is well-defined. Therefore, the odd $N$ case, will have a different charge parity and cannot belong to the same ground state manifold \cite{Read2000,Oshikawa2007,Iadecola2019}.

\emph{Heisenberg algebra} -- In this section we probe the topology of the SO(N) model using operator formalism \cite{Oshikawa2007,Iadecola2019}. We limit ourselves to the even $N$ case, and defer the more subtle odd $N$ case to future \cite{Ljepoja2024b}.  We use the fact that critical SO(N) model can be regarded as a collection of $N$ Ising \emph{layers}, which for even $N$ can be paired with each other. To simplify the matters, we analyze the problem when the size of the system is sent to infinity so that the topological ground state degeneracy is exact.

Consider a set of $N=2M$ dimension-1/2 fermions $\chi^a$ with $a=1\dots N=2M$  which transform into each other as a vectorial representation of SO(N), i.e. $\chi^a\to R^a_{\hh b}\chi^b$. They can be bosonized by introducing $M$ bosons $\phi_r$ so that $\chi^{2r-1}=\sqrt{2}\cos\phi_r$ and $\chi^{2r}=\sqrt{2}\sin\phi_r$. For $N\ge 2$, $\pi_1[SO(N)]=\bb Z_2$ meaning that SO(N) has a double cover, known as Spin(N). For example for SO(5) this is Spin(5)$\sim$ SP(4). This cover group admits a spinorial representation, $2^M$ new operators $\omega^{\alpha}\equiv e^{\frac{i}{2}\vec\alpha\cdot\vec\phi}$ defined in terms of $\phi_r$ and elements $\alpha_r=\pm 1$. These operators have the dimension $[\omega^{\alpha}]=N/16$ equivalent to the product of $N$ chiral twist operators from the $N$ Ising sectors. For any $r$ one can find $\alpha$, $\beta$ such that $\phi_r=\frac{1}{2}(\vec\beta-\vec\alpha)\cdot\vec\phi$. Therefore, $\chi^a$ operators can be expressed in terms of $\omega^\alpha$ operators, $\chi^a=\omega{^\alpha}\dg {\cal T}^a_{\alpha\beta}\omega^\beta$,   with $\omega$ operators transforming as the spinorial (fundamental) representation $\omega^\alpha\to U^{\alpha}_{\hh\beta}\omega^\beta$ of Spin(N), i.e. $U\dg [{\cal T}^a]U=R^a_b{\cal T}^b$. Since $\omega{^\alpha}\dg=\omega^{-\alpha}$, $\chi$ is invariant under $\omega^\alpha\to-\omega^\alpha$ only and therefore, each element of SO(N) corresponds to only two elements in Spin(N). In CFT these are the spinorial primaries of the SO(N) WZW model \cite{DiFrancesco1999}. The commutation relation of $\chi$ and $\omega$ can be found from the Baker-Campbell-Hausdorff formula \cite{SM}. Defining the SO(N) symmetric mutually commuting operators on arbitrary wire $j$
\be
\hspace{-.07cm}\Gamma_y^1\equiv i\sum_a\chi^a_j(0)\chi^a_j(L), \hspace{.3cm} \eps_j(x)\equiv i\sum_a{\bar \chi}^a_j(x) \chi_{j+1}^a(x)\hspace{-.12cm}
\ee
and a family of tensorial operators
\be
\Gamma_x^{1/2}\equiv\sum_{\{\alpha,\beta\}}{\cal N}_{\{\alpha,\beta\}}\prod_j\bar\omega{^{-\alpha_j}_j}(x_j)\omega^{\beta_j}_{j+1}(x_j),
\ee
it follows that irrespective of the tensor ${\cal N}_{\{\alpha,\beta\}}$, we have $\Gamma_x^{1/2}\Gamma_y^{1}=-\Gamma_y^{1}\Gamma_x^{1/2}$ and $\Gamma_x^{1/2}\eps_j(x)=\eps_j(x)\Gamma_x^{1/2}$. It can be shown \cite{SM} that $\Gamma_x^{1/2}$ introduces a discontinuity $\sgn(x-x_j)$ into each $\chi_j^a(x)$ or $\bar\chi^a_j(x)$, and at the same time shuffling $\chi^{2r}\lr\chi^{2r-1}$. While the precise position of sign change is gauge-dependent and can be moved via a gauge transformation, the boundary condition over these Majorana fields is changed from anti-periodic to periodic or vice versa ($\pi$-flux). Likewise, we define the operators
\bea
\Gamma_x^1\equiv\prod_j\eps_j(x_j), \qquad
\Gamma_y^{1/2}&\equiv&\sum_{\alpha}\omega^{-\alpha}_j(0)\omega^\alpha_{j}(L)
\eea
It can be shown that $\Gamma_y^{1/2}\chi^a_j(x)\Gamma_y^{1/2}=-\chi^a_j(x)$ and these operators implement the KW symmetry. Consequently,  $\Gamma_y^{1/2}\Gamma_x^{1}=-\Gamma_x^{1}\Gamma_y^{1/2}$ and $\Gamma_y^{1/2}\Gamma_y^{1}=\Gamma_y^{1}\Gamma_y^{1/2}$. All $\Gamma$ operators commute with the Hamiltonian. $\Gamma_{x,y}^{1}$ can be used to label the ground state $\ket{\Gamma_x^{1}\Gamma_y^1}$, while $\Gamma_{x,y}^{1/2}$ due to their anti-commutation relation $\{\Gamma^{1/2}_x,\Gamma_y^1\}=\{\Gamma^{1/2}_y,\Gamma_x^1\}=0$ can be used to switch between these states. Therefore, for $N$ even, the SO(N) model \pref{eqSON} has GSD of 4, in agreement with the number of primaries of the CFT \cite{Francesco1997,Ftnote3}.

\emph{Conclusion} -- Using CWC, we have studied the PM regime of a TRS broken 2CKL with spin-1/2 SU(2) local moments. In presence of the PHS, the original  SO(8) symmetry is reduced to SO(5)$\times$Ising, which are gauge-coupled, and a decoupled SU(2)$_1$ sectors. The inter-wire coupling increases the symmetry to SO(6)$\otimes$SU(2)$_1$. We have analyzed the GSD of this gapped phase on a torus, showing that it is eight-fold degenerate, 4 from SO(6) and 2 from SU(2)$_1$. The assumption of PHS has led to an abelian topological order. When PHS is broken it is possible to write down a SO(3) model, with a non-abelian ground state \cite{Iadecola2019}. We defer the PHS analysis to \cite{Ljepoja2024b}.

\emph{Outlook} -- There are many ways in which these results can be extended; the WKC and PM regimes share the same edge states, despite the different bulk physics. In both phases, $\phi_c$, $\phi_f$ and $\phi_J$ bosons are pinned to the neighbors, whereas in PM regime the remaining bosons $\phi_\gamma$ and $\phi_{sf}$ are passivated partly inside the wire and partly by the neighbors. This raises questions about bulk-boundary correspondence. Lastly, our picture of FM phase as the proliferation of bosonic SP(4) spinorial defects rather than SO(5), points toward order fractionalization \cite{ofc}, a notion related to deconfined criticality \cite{Senthil2004,Huang2021,Huang2023} but extended over an entire phase. This is a fascinating direction to be explored in the future.

\bibliography{Anyons2}

\begin{thebibliography}{87}
\bibitem [{\citenamefont {{There are eight states, but $i\eta^a\eta^b$ does not
  change the fermionic parity, only shuffling those with the same parity.
  $\eta^a$ operators do change the parity, but they are not
  gauge-invariant}}()}]{Ftnote1}%
  \BibitemOpen
  \bibfield  {author} {\bibinfo {author} {\bibnamefont {{There are eight
  states, but $i\eta^a\eta^b$ does not change the fermionic parity, only
  shuffling those with the same parity. $\eta^a$ operators do change the
  parity, but they are not gauge-invariant}}},\ }\href@noop {} {}\BibitemShut
  {NoStop}%
\bibitem [{\citenamefont {{$(\Gamma_{x,y}^1)^2=1$ and therefore we have four
  states $\ket{++}$ ,$\ket{+-}$, $\ket{-+}$ and $\ket{--}$}}()}]{Ftnote3}%
  \BibitemOpen
  \bibfield  {author} {\bibinfo {author} {\bibnamefont {{$(\Gamma_{x,y}^1)^2=1$
  and therefore we have four states $\ket{++}$ ,$\ket{+-}$, $\ket{-+}$ and
  $\ket{--}$}}},\ }\href@noop {} {}\BibitemShut {NoStop}%
\end{thebibliography}
\newpage

\bw

\section*{Supplementary Materials}

\subsection*{RG analysis}
Interacting Hamiltonian density for this model is:
\begin{equation}
\begin{split}
H(z, \bar{z}) = H_{0} (z, \bar{z})+ \lambda_{0}\sum_{j} \sum_{a} \bar{{\cal O}}^{a}_{j} (\bar{z}) {\cal O}^{a}_{j+1}(z) + \lambda_{I} \sum_{j} \sum_{a, b} \bar{{\cal I}}^{ab}_{j}(\bar{z}) {\cal I}^{ab}_{j+1}(z)
\end{split}
\end{equation}
Where we have defined:
\begin{equation}
\begin{split}
     & {\cal O}^{a}_{j}(z) = i \bar{\gamma}_{j}(\bar{z}) \chi^{a}_{j} (z)\\
     & {\cal I}^{ab}_{j}(z) = i \chi^{a}_{j}(z) \chi^{b}_{j}
\end{split}
\end{equation}
Operators $\gamma$ and $\chi$ are Majorana fermions and they satisfy:
\begin{equation}
    \gamma(z_1)\gamma(z_2) \equiv \frac{1}{z_1-z_2}, \qquad \qquad \chi^{a}(z_1)\chi^{b}(z_2) \equiv \frac{\delta^{ab}}{z_1-z_2}
\end{equation}
Above OPE's are used to calculate the OPE's between $\cal O$ and $\cal I$ operators, which we shall need for the RG flow.
\begin{equation}
\begin{split}
    {\cal O}^{a}(z_{1}) {\cal O}^{b}(z_{2}) & = [ i \bar{\gamma}_{j}(\bar{z}_1) \chi^{a}_{j} (z_1)] \, [i \bar{\gamma}_{j}(\bar{z}_2) \chi^{b}_{j} (z_2)]\\
    & = \frac{\chi^{a}_{j} (z_1) \chi^{b}_{j} (z_2)}{(z_1 - z_2)} +\frac{\delta^{ab}}{(z_1 - z_2)^2}\\
    & = \frac{-i I^{ab}(z_{12})}{\Delta z} +\frac{\delta^{ab}}{(\Delta z)^2}
\end{split}
\end{equation}
And accordingly for the right sector. The first term comes from the contraction of $\bar{\gamma}$'s only while the second term is when we contract both $\bar{\gamma}$'s with itself and $\chi$'s with itself. Contraction of just $\chi$'s does not give any contribution since the resulting product of $\bar{\gamma}(\bar{z}_1)\bar{\gamma}(\bar{z}_2)$ would be zero in the limit $z_1 \to z_2$. Also we have defined $I^{ab}_{z_{12}} = i \chi^{a}_{j} (z_1) \chi^{b}_{j} (z_2)$. Next, we look at the OPE of ${\cal O}$ and ${\cal I}$:
\begin{equation}
\begin{split}
    {\cal O}_{j}^{a}(z_{1}) {\cal I}_{j}^{bc}(z_{2}) & = [ i \bar{\gamma}_{j}(\bar{z}_1) \chi^{a}_{j} (z_1)] \, [i \chi^{b}_{j}(z_2) \chi^{c}_{j} (z_2)]\\
    & = -\frac{\bar{\gamma}_{j} (\bar{z}_1) \chi^{c}_{j} (z_2) \delta^{ab}}{(z_1 - z_2)} +\frac{\bar{\gamma}_{j} (\bar{z}_1) \chi^{b}_{j} (z_2)\delta^{ac}}{(z_1 - z_2)}\\
    & = \frac{i {\cal O}^{c}(z_{12}) \delta^{ab}}{\Delta z} - \frac{i {\cal O}^{b}(z_{12})\delta^{ac}}{\Delta z}
\end{split}
\end{equation}
Where, the first term comes from the contraction of $\chi^{a}$ and $\chi^{b}$ amongst each other and the second is from a similar contraction but between $\chi^{a}$ and $\chi^{c}$. There are no other contractions that can be made.
\begin{equation}
\begin{split}
    {\cal I}_{j}^{ab}(z_{1}) {\cal I}_{j}^{cd}(z_{2}) & = [ i \chi^{a}_{j}(z_1) \chi^{b}_{j} (z_1)] \, [i \chi^{c}_{j}(z_2) \chi^{d}_{j} (z_2)]\\
    & = \frac{\chi_{j}^{b} (z_1) \chi^{d}_{j} (z_2) \delta^{ac}}{(z_1 - z_2)} - \frac{\chi_{j}^{b} (z_1) \chi^{c}_{j} (z_2) \delta^{ad}}{(z_1 - z_2)} - \frac{\chi_{j}^{a} (z_1) \chi^{d}_{j} (z_2) \delta^{bc}}{(z_1 - z_2)} + \frac{\chi_{j}^{a} (z_1) \chi^{c}_{j} (z_2) \delta^{bd}}{(z_1 - z_2)} +\frac{\delta^{ac} \delta^{bd}}{(z_1 - z_2)^2} - \frac{\delta^{ad} \delta^{bc}}{(z_1 - z_2)^2}\\
    & = -\frac{i {\cal I}^{bd}(z_{12}) \delta^{ac}}{\Delta z} + \frac{i {\cal I}^{bc}(z_{12})\delta^{ad}}{\Delta z} + \frac{i {\cal I}^{ad}(z_{12})\delta^{bc}}{\Delta z} - \frac{i {\cal I}^{ac}(z_{12})\delta^{bd}}{\Delta z} +\frac{\delta^{ac} \delta^{bd}}{(\Delta z)^2} - \frac{\delta^{ad} \delta^{bc}}{(\Delta z)^2}
\end{split}
\end{equation}
Most of these terms will cancel once we multiply with the right sector OPE and do the summation over the indices.

Armed with these OPE's now we can look at the RG flow of the Hamiltonian. Using the S matrix expansion we have:
\begin{equation}
\begin{split}
S=T_{\tau} e^{-\int dz d\bar{z} H (z, \bar{z})} \approx 1 - T_{\tau}\int d^2z \, H(z, \bar{z}) + \frac{1}{2}T_{\tau}\int d^2 z_{1} \int d^2 z_{2} \, H (z_{1}, \bar{z}_{1}) \, H (z_{2}, \bar{z}_{2}) + \ldots
\end{split}
\end{equation}
Where we have written $d^{2} = dz d\bar{z}$ and $T_{\tau}$ is radial ordering operator. The second order term (which we are interested in) will be of the form:
\begin{equation}
\label{eq:master}
\begin{split}
\frac{1}{2}T_{\tau}\int d^2 z_{1} \int d^2 z_{2} & \, \bigg [ \lambda_{0}\sum_{j} \sum_{a} \bar{{\cal O}}^{a}_{j} (\bar{z}_1) {\cal O}^{a}_{j+1}(z_1) + \lambda_{I} \sum_{j} \sum_{a, b} \bar{{\cal I}}^{ab}_{j}(\bar{z}_1) {\cal I}^{ab}_{j+1}(z_1) \bigg ] \\ 
& \qquad \times \bigg [ \lambda_{0}\sum_{j} \sum_{a^{\prime}} \bar{{\cal O}}^{a^{\prime}}_{j} (\bar{z}_2) {\cal O}^{a^{\prime}}_{j+1}(z_2) + \lambda_{I} \sum_{j} \sum_{a^{\prime}, b^{\prime}} \bar{{\cal I}}^{a^{\prime}b^{\prime}}_{j}(\bar{z}_2) {\cal I}^{a^{\prime} b^{\prime}}_{j+1}(z_2) \bigg ]
\end{split}
\end{equation}
The idea is to integrate out the high energy degrees of freedom by changing the regularization from $a \to a+ da$ which comes out as a multiplicative contribution after integrating over that extended area. Term by term we have:
\begin{equation*}
\begin{split}
\frac{1}{2}T_{\tau} \lambda^{2}_{0} \sum_{j} \sum_{a, a^{\prime}}\int d^2 z_{1} \int d^2 z_{2}  \, \bigg [ & \bar{{\cal O}}^{a}_{j} (\bar{z}_1) {\cal O}^{a}_{j+1}(z_1)\bigg ] \bigg [ \bar{{\cal O}}^{a^{\prime}}_{j} (\bar{z}_2) {\cal O}^{a^{\prime}}_{j+1}(z_2)\bigg ] \\
= \frac{1}{2}T_{\tau} \lambda^{2}_{0} \sum_{j} \sum_{a, a^{\prime}}\int d^2 z_{1} \int d^2 z_{2} \bigg [ & \, \bigg (\frac{-i \bar{{\cal I}}^{aa^{\prime}}_{j}(\bar{z}_{12})}{\Delta \bar{z}} +\frac{\delta^{aa^{\prime}}}{(\Delta \bar{z})^2} \bigg ) {\cal O}^{a}_{j+1} (z_1) {\cal O}^{a^{\prime}}_{j+1}(z_2)+ \\
& + \bigg ( \frac{-i {\cal I}_{j+1}^{aa^{\prime}}(z_{12})}{\Delta z} +\frac{\delta^{aa^{\prime}}}{(\Delta z)^2} \bigg) \bar{{\cal O}}^{a}_{j} (\bar{z}_1) \bar{{\cal O}}^{a^{\prime}}_{j}(\bar{z}_2) +\\
& + \bigg (\frac{-i \bar{{\cal I}}^{aa^{\prime}}_{j}(\bar{z}_{12})}{\Delta \bar{z}} +\frac{\delta^{aa^{\prime}}}{(\Delta \bar{z})^2} \bigg )  \bigg ( \frac{-i {\cal I}_{j+1}^{aa^{\prime}}(z_{12})}{\Delta z} +\frac{\delta^{aa^{\prime}}}{(\Delta z)^2} \bigg) \bigg ] \\
\end{split}
\end{equation*}
The first two terms in the integrand will be zero since they will involve product of the form $\gamma(z_1) \gamma(z_2)$. Some of the contributions in the last term in the above expression will involve product of Kronecker delta, demanding that $a=a^{\prime}$ which will make the current $I^{aa} \to 0$ in the low energy limit. Taking care of all these cancellations we are left only with:
\begin{equation*}
-\frac{1}{2}T_{\tau} \lambda^{2}_{0} \sum_{j} \sum_{a, a^{\prime}}\int d^2 z_{12} \, \bar{{\cal I}}^{aa^{\prime}}_{j}(\bar{z}_{12}) {\cal I}_{j+1}^{aa^{\prime}}(z_{12}) \int_{a}^{a+da} \frac{d\Delta z}{\Delta z} \int_{a}^{a+da} \frac{d\Delta \bar{z}}{\Delta \bar{z}}  + T_{\tau} \lambda^{2}_{0} \sum_{j} \sum_{a}\int d^2 z_{12} \int_{a}^{a+da} \frac{d \Delta z}{(\Delta z)^2} \int_{a}^{a+da} \frac{d \Delta \bar{z}}{(\Delta \bar{z})^2}
\end{equation*}
Integrals are given by:
\begin{equation}
\begin{split}
\int_{a}^{a+da} \frac{d\Delta z}{\Delta z} \int_{a}^{a+da} \frac{d\Delta \bar{z}}{\Delta \bar{z}} = 2 \ln(1+\frac{da}{a}) \approx 2\frac{da}{a} = 2 dl\\
\int_{a}^{a+da} \frac{d \Delta z}{(\Delta z)^2} \int_{a}^{a+da} \frac{d \Delta \bar{z}}{(\Delta \bar{z})^2} = \frac{da^2}{a^2 (a+da)^2} \to 0
\end{split}
\end{equation}
Therefore we finally have:
\begin{equation*}
\begin{split}
-T_{\tau}  dl \lambda^{2}_{0} \sum_{j} \sum_{a, a^{\prime}}\int d^2 z_{12} \, \bar{{\cal I}}^{aa^{\prime}}_{j}(\bar{z}_{12}) {\cal I}_{j+1}^{aa^{\prime}}(z_{12})  = -2 T_{\tau} dl \lambda^{2}_{0} \sum_{j} \sum_{a < a^{\prime}}\int d^2 z_{12} \, \bar{{\cal I}}^{aa^{\prime}}_{j}(\bar{z}_{12}) I_{j+1}^{a a^{\prime}}(z_{12})
\end{split}
\end{equation*}
Note that we have introduced a restricted sum which resulted with in a overall factor of $2$. The flow is:
\begin{equation}
    \lambda_{I}^{\prime} = \lambda_{I} + 2dl \lambda_{0}^2 \longrightarrow \frac{d \lambda_{I}}{dl} =  2\lambda^{2}_{0}
\end{equation}
Where the extra minus sign comes from the minus sign in the exponential of the S matrix.
There will be other terms from Eq.~(\ref{eq:master}) that contribute to the flow of $\lambda_{I}$. Namely:
\begin{equation}
\begin{split}
\frac{1}{2}T_{\tau} \lambda_{I}^2 \sum_{j} \sum_{a < b} \sum_{a^\prime < b^\prime}\int d^2 z_{1} \int d^2 z_{2} & \, \bigg [ \bar{{\cal I}}^{ab}_{j}(\bar{z}_1) I^{ab}_{j+1}(z_1) \bigg ]\times \bigg [ \bar{{\cal I}}^{a^{\prime}b^{\prime}}_{j}(\bar{z}_2) {\cal I}^{a^{\prime} b^{\prime}}_{j+1}(z_2) \bigg ] \\
= \frac{1}{2} T_{\tau} \lambda_{I}^2 \sum_{j} \sum_{a < b} \sum_{a^\prime < b^\prime} \int d^2 z_{1} \int d^2 z_{2} & \, \bigg [ -\frac{i \bar{I}^{bb^\prime}_{j}(\bar{z}_{12}) \delta^{aa^\prime}}{\Delta \bar{z}} + \frac{i \bar{{\cal I}}^{ba^\prime}_{j}(\bar{z}_{12})\delta^{ab^\prime}}{\Delta \bar{z}} + \frac{i \bar{{\cal I}}^{a b^\prime}_{j}(\bar{z}_{12})\delta^{ba^\prime}}{\Delta \bar{z}} - \frac{i \bar{{\cal I}}^{aa^\prime}_{j}(\bar{z}_{12})\delta^{bb^\prime}}{\Delta \bar{z}} \bigg ]\\
& \qquad \times \bigg [-\frac{i {\cal I}^{bb^\prime}_{j+1}(z_{12}) \delta^{aa^\prime}}{\Delta z} + \frac{i {\cal I}^{ba^\prime}_{j+1}(z_{12})\delta^{ab^\prime}}{\Delta z} + \frac{i {\cal I}^{ab^\prime}_{j+1}(z_{12})\delta^{ba^\prime}}{\Delta z} - \frac{i {\cal I}^{aa^\prime}_{j+1}(z_{12})\delta^{bb^\prime}}{\Delta z} \bigg ]
\end{split}
\end{equation}
In obtaining the expression in the second line we have neglected the energy shift terms (terms of order $1/(\Delta z)^2$) which go to zero once we set $\frac{da}{a} \to 0$, and also we have not written the contractions between just one pair of currents since that, as we have seen in the derivation of the previous terms will be zero. Looking at the above equations we can see (due to the condition $a<b$ and $a^\prime<b^\prime$) that all the cross products (one that have two different Kronecker deltas) are zero because the Kronecker delta cannot be simultaneously satisfied.  We can do the integral in $\Delta z$ and we obtain the same result as before (it gives a factor of $2 dl$) and rewrite the condition on the sums using Heaviside theta functions:
\begin{equation}
\begin{split}
-T_{\tau} \lambda_{I}^2 dl\sum_{j} \sum_{a, b} \sum_{a^\prime, b^\prime} \int d^2 z_{12} \, \bar{{\cal I}}^{bb^\prime}_{j}(\bar{z}_{12}) {\cal I}^{bb^\prime}_{j+1}(z_{12}) \delta^{aa^\prime} & \, \bigg [ \theta(b-a)\theta(b^\prime - a^\prime) + \theta(b-a)\theta(a^\prime - b^\prime) \\
& + \theta(a-b)\theta(b^\prime - a^\prime)  + \theta(a-b)\theta(a^\prime - b^\prime) \bigg ] 
\end{split}
\end{equation}
where the sums are now unrestricted and go over all values. If we use the Kronecker delta now to kill the sum over $a^{\prime}$ we are left with current current interaction muptiplied by a sum:
\begin{equation}
\begin{split}
\sum_{a}\bigg [ \theta(b-a)\theta(b^\prime - a) + \theta(b-a)\theta(a - b^\prime)+ \theta(a-b)\theta(b^\prime - a)  + \theta(a-b)\theta(a - b^\prime) \bigg ]  = \sum_{a}1 = N
\end{split}
\end{equation}
Where $N$ is the number of flavours of $\chi$ and we have used the property of the step function that $\theta(x)+\theta(-x) = 1$. So now finally we have:
\begin{equation}
\begin{split}
-T_{\tau} \lambda_{I}^2 dl N \sum_{j} \sum_{b, b^\prime} \int d^2 z_{12} \, \bar{I}^{bb^\prime}_{j}(\bar{z}_{12}) I^{bb^\prime}_{j+1}(z_{12}) = -T_{\tau} 2\lambda_{I}^2 dl N \sum_{j}  \sum_{b < b^\prime} \int d^2 z_{12} \, \bar{{\cal I}}^{bb^\prime}_{j}(\bar{z}_{12}) {\cal I}^{b b^\prime }_{j+1}(z_{12})
\end{split}
\end{equation}
This term adds to the flow of the $\lambda_{I}$ and together with the $\lambda_{0}^2$ gives:
\begin{equation}
    \lambda_{I}^{\prime} = \lambda_{I} + 2dl \lambda_{0}^2 + 2N dl \lambda_{I}^2 \longrightarrow \frac{d \lambda_{I}}{dl} =  2\lambda^{2}_{0} + 2N\lambda^{2}_{I}
\end{equation}
We have the flow for $\lambda_{I}$, now we can look at the flow of $\lambda_{0}$. It is obtained from contractions between ${\cal O}$'s and $I$'s in eq.~(\ref{eq:master}). We have:
\begin{equation}
\begin{split}
2 \times T_{\tau}\frac{1}{2}\lambda_{0} \lambda_{I} \sum_{j} \sum_{a} \sum_{a^{\prime} < b^{\prime}}\int d^2 z_{1} \int d^2 z_{2} & \, \bigg [\bar{{\cal O}}^{a}_{j} (\bar{z}_1) {\cal O}^{a}_{j+1}(z_1)  \bigg ]  \times \bigg [\bar{{\cal I}}^{a^{\prime}b^{\prime}}_{j}(\bar{z}_2) {\cal I}^{a^{\prime} b^{\prime}}_{j+1}(z_2) \bigg ]
\end{split}
\end{equation}
Where the extra factor of two comes from the fact that the ${\cal O}$ operators and currents commute, being composed of a pair of majoranas. Using the OPE of ${\cal O}\cdot I$ we can rewrite the above expression:
\begin{equation}
\begin{split}
T_{\tau} \lambda_{0} \lambda_{I} \sum_{j} \sum_{a} \sum_{a^{\prime} < b^{\prime}}\int d^2 z_{1} \int d^2 z_{2} & \, \bigg [\frac{i \bar{{\cal O}}^{b^\prime}_{j}(\bar{z}_{12}) \delta^{aa^{\prime}}}{\Delta \bar{z}} - \frac{i \bar{{\cal O}}^{a^{\prime}}_{j}(\bar{z}_{12})\delta^{ab^{\prime}}}{\Delta \bar{z}}  \bigg ]  \times \bigg [\frac{i {\cal O}^{b^{\prime}}_{j+1}(z_{12})\delta^{aa^{\prime}}}{\Delta z} - \frac{i {\cal O}^{a^{\prime}}_{j+1}(z_{12}) \delta^{ab^{\prime}}}{\Delta z} \bigg ]
\end{split}
\end{equation}
In the above expression the cross-products do not survive since $\delta^{aa^\prime}\delta^{ab^\prime} = \delta^{a^\prime b^\prime}$ can ever be satisfied and it is always zero. Therefore, after integrating out fast modes (doing th eintegral in $\Delta z$) we arrive at:
\begin{equation}
\begin{split}
-T_{\tau} 2 \lambda_{0} \lambda_{I} dl \sum_{j} \sum_{a} \sum_{a^{\prime} < b^{\prime}}\int d^2 z_{11}  & \, \bigg [\bar{{\cal O}}^{b^\prime}_{j}(\bar{z}_{12}) {\cal O}^{b^{\prime}}_{j+1}(z_{12}) \delta^{aa^{\prime}}  +  \bar{{\cal O}}^{a^{\prime}}_{j}(\bar{z}_{12}) {\cal O}^{a^{\prime}}_{j+1}(z_{12}) \delta^{ab^{\prime}} \bigg ]
\end{split}
\end{equation}
Like in the previous calculation we can introduce step functions and than freely exchange dummy indices so we get:
\begin{equation}
\begin{split}
-T_{\tau} 2 \lambda_{0} \lambda_{I} dl \sum_{j} \sum_{a} \sum_{a^{\prime}, b^{\prime}}\int d^2 z_{11} & \, \bar{{\cal O}}^{b^\prime}_{j}(\bar{z}_{12}) {\cal O}^{b^{\prime}}_{j+1}(z_{12}) \delta^{aa^{\prime}}  \bigg [ \theta(b^\prime - a^\prime) + \theta(a^\prime - b^\prime) \bigg ] \\
= -T_{\tau} 2 \lambda_{0} \lambda_{I} dl \sum_{j} \sum_{a} \sum_{a^{\prime}, b^{\prime}}\int d^2 z_{11} & \, \bar{{\cal O}}^{b^\prime}_{j}(\bar{z}_{12}) {\cal O}^{b^{\prime}}_{j+1}(z_{12}) \delta^{aa^{\prime}}
\end{split}
\end{equation}
The sum over $a$ and $a^{\prime}$ in the integrand is nothing but the number of flavours of the $\chi$'s as can be seen:
\begin{equation}
\begin{split}
 \sum_{a} \sum_{a^{\prime}}\delta^{aa^{\prime}} = N
\end{split}
\end{equation}
So the renormalized $\lambda_{0}$ becomes:
\begin{equation}
\begin{split}
-T_{\tau} 2 \lambda_{0} \lambda_{I} N dl \sum_{j} \sum_{b^{\prime}}\int d^2 z_{11} & \, \bar{{\cal O}}^{b^\prime}_{j}(\bar{z}_{12}) {\cal O}^{b^{\prime}}_{j+1}(z_{12}) 
\end{split}
\end{equation}
giving the flow of $\lambda_{0}$ as:
\begin{equation}
    \frac{d \lambda_{0}}{dl} = 2N \lambda_{0} \lambda_{I}
\end{equation}

\subsection*{Identification of the Tolouse point}
Inter-wire Hamilonian density describes the Kondo exchange between the $SU(2)_{2}$ fermion currents  and $SU(2)_{1}$ spin current and is given by:
\begin{equation*}
\begin{split}
    &H=H_{1}+H_{2}\\
    &H_{1} = \frac{\pi v_{1}}{2} {\cal \vec{J}}_{R}\cdot {\cal \vec{J}}_{R} + \frac{2 \pi v_{0}}{3} {\cal \vec{S}}_{L}\cdot {\cal \vec{S}}_{L} +g_{z} {\cal J}^{z}_{R}{\cal S^{z}}_{L}+\frac{g_{\perp}}{2}({\cal J}_{R}^{+}{\cal S}^{-}_{L}+H.c)\\
    &H_{2} = H_{1}(L \leftrightarrow R)
\end{split}
\end{equation*}
Where $\cal \vec{J}$ and $\cal \vec{S}$ are $SU(2)_{2}$ and $SU(2)_{1}$ currents respectively. The solution of the above model is known and It is based on the mapping to the Majorana fermions as well as on the existence of the Toulouse-like, exactly solvable point.  As it was mentioned in our main text, due to the presence of the Toulouse point the original $SU(2)$ symmetry of the model is restored. The $SU(2)_{1}$ currents can be readily bosonized by introducing a massless bosonic field $\phi$:
\begin{equation}
\begin{split}
   &{\cal S}^{z}_{R} \sim \partial_{x} \bar{\phi}_{m}, \qquad \qquad {\cal S}^{\pm}_{R} \sim e^{\mp \sqrt{2} \bar{\phi}_{m}}\\
   &{\cal S}^{z}_{L} \sim \partial_{x} \phi_{m}, \qquad \qquad {\cal S}^{\pm}_{L} \sim e^{\mp \sqrt{2} \bar{\phi}_{m}}
\end{split}
\end{equation}
These currents can be shown to satisfy the necessary OPE's of the $SU(2)_{1}$ WZNW models. On the other hand the components of the $SU(2)_2$ current, $\cal \vec{J}$ can be written in terms of Majorana fermions:
\begin{equation*}
    {\cal I}^{c}_{R} = i \epsilon^{abc} \bar{\chi}_{a} \bar{\chi}_{b}
\end{equation*}
And likewise for the $L$ currents. In the paper those majoranas are identified as $\chi_{1} = \chi^{\prime}_{s}$, $\chi_{2} = \chi^{\prime \prime}_{s}$ and $\chi_{3} = \chi^{\prime}_{sf}$. One can define a Dirac fermion from the  $\chi^{\prime}_{s}$ and $\chi^{\prime \prime}_{s}$ Majoranas as:
\begin{equation}
\begin{split}
    &\bar{\psi}_{s} = \frac{\bar{\chi}^{\prime}_{s} + i\bar{\chi}^{\prime \prime}_{s}}{\sqrt{2}} \sim \kappa e^{i \bar{\phi}_s}\\
    & \psi_{s} = \frac{\chi^{\prime}_{s} + i\chi^{\prime \prime}_{s}}{\sqrt{2}} \sim \kappa e^{i \phi_s}\\
\end{split}
\end{equation}
This enables us to bosonize the $SU(2)_{2}$ currents as:
\begin{equation}
\begin{split}
    & {\cal J}^{z}_{R} = i \bar{\chi}^{\prime}_{s}\bar{\chi}^{\prime\prime}_{s} = :\bar{\psi}^{\dagger}_{s} \bar{\psi}_{s}: \sim  \partial_{x} \bar{\phi}_{s}, \qquad \qquad {\cal J}_{R}^{\pm} = i\bar{\chi}^{\prime \prime}_{s}\bar{\chi}^{\prime}_{sf} \mp \bar{\chi}^{\prime}_{sf} \bar{\chi}^{\prime}_{s} \sim \chi^{\prime}_{sf}\kappa e^{\mp i \bar{\phi}_{s}}\\
    & {\cal J}^{z}_{L} = i \chi^{\prime}_{s}\chi^{\prime\prime}_{s} = :\psi^{\dagger}_{s} \psi_{s}: \sim  \partial_{x} \phi_{s}, \qquad \qquad {\cal J}_{R}^{\pm} = i\chi^{\prime \prime}_{s}\chi^{\prime}_{sf} \mp \chi^{\prime}_{sf} \chi^{\prime}_{s} \sim \chi^{\prime}_{sf}\kappa e^{\mp i \phi_{s}}
\end{split}
\end{equation}
Using these bosonization identities we arrive at the bosonized form of the Hamiltonian:
\begin{equation}
\begin{split}
    & H_{1} \sim \frac{2\pi v_{0}}{3}(\partial_{x} \phi_{m})^2+\frac{\pi v_{1}}{2}(\partial_{x} \bar{\phi}_{s})^2 - i\frac{\pi}{4v_1} \bar{\chi}^{\prime}_{sf}\partial_{x} \bar{\chi}^{\prime}_{sf} + g_{z} (\partial_{x}\phi_{m}) (\partial_{x}\bar{\phi}_{s}) + \frac{g_{\perp}}{2} \bar{\chi}^{\prime}_{sf} \kappa \cos (\bar{\phi_{s}} + \sqrt{2} \phi_{m}) \\
    & H_{2} \sim \frac{2\pi v_{0}}{3}(\partial_{x} \bar{\phi}_{m})^2+\frac{\pi v_{1}}{2}(\partial_{x} \phi_{s})^2 - i\frac{\pi}{4v_1} \chi_{sf}^{\prime} \partial_{x} \chi^{\prime}_{sf} + g_{z} (\partial_{x}\bar{\phi}_{m}) (\partial_{x}\phi_{s}) + \frac{g_{\perp}}{2} \chi^{\prime}_{sf} \kappa \cos (\phi_{s} + \sqrt{2} \bar{\phi_{m}}) \\
\end{split}
\end{equation}
The $H_{2}$ is obtained in the same way as $H_{1}$ just exchanging L and R,
In the above expression we can use a transformation to eliminate the cross terms:
\begin{equation}
    \begin{pmatrix}
        \phi_{m} \\
        \bar{\phi}_{s}
    \end{pmatrix}=
    \begin{pmatrix}
        \cosh(\alpha) & \sinh(\alpha)\\
        \sinh(\alpha) & \cosh(\alpha)
    \end{pmatrix}
    \begin{pmatrix}
        \phi_{\gamma} \\
        \bar{\phi}_{J}
    \end{pmatrix}, \qquad \qquad
    \begin{pmatrix}
        \phi_{s} \\
        \bar{\phi}_{m}
    \end{pmatrix}=
    \begin{pmatrix}
        \cosh(\alpha) & \sinh(\alpha)\\
        \sinh(\alpha) & \cosh(\alpha)
    \end{pmatrix}
    \begin{pmatrix}
        \phi_{J} \\
        \bar{\phi}_{\gamma}
    \end{pmatrix}
\end{equation}
which transforms the cross terms as:
\begin{equation}
    \bigg [g_z (\sinh^2(\alpha)+\cosh^2(\alpha)) + 2\pi(\frac{2v_{0}}{3}+\frac{v_1}{2})\sinh(\alpha)\cosh(\alpha) \bigg ] (\partial_x \phi_{\gamma})(\partial_{x} \bar{\phi}_{J}) 
\end{equation}
It is obvious that this terms will be equal to zero if we choose $\alpha$ such that:
\begin{equation}
    \frac{2\sinh(\alpha)\cosh(\alpha)}{\sinh^2(\alpha)+\cosh^2(\alpha)} \equiv \tanh(2\alpha) = -\frac{g_z}{4 \pi (\frac{v_0}{3}+\frac{v_1}{4})}
\end{equation}
On the other hand such transformation changes the fields under the cosine in the Hamiltoian in the following way:
\begin{equation}
\begin{split}
&\bar{\phi}_{s} + \sqrt{2} \phi_{m} \to \big[(\sqrt{2}\cosh(\alpha)+\sinh(\alpha))\phi_{\gamma} + (\cosh(\alpha)+\sqrt{2}\sinh(\alpha))\bar{\phi}_{J}\big]\\
&\phi_{s} + \sqrt{2} \bar{\phi}_{m} \to \big[(\sqrt{2}\cosh(\alpha)+\sinh(\alpha))\bar{\phi}_{\gamma} + (\cosh(\alpha)+\sqrt{2}\sinh(\alpha))\phi_{J}\big]\\
\end{split}
\end{equation}
Suitably choosing $\tanh(\alpha) = -\frac{1}{\sqrt{2}}$ we can see that we obtain a transformation that maps:
\begin{equation}
\begin{split}
&\bar{\phi}_{s} + \sqrt{2} \phi_{m} \to \phi_{\gamma} \\
&\phi_{s} + \sqrt{2} \bar{\phi}_{m} \to \bar{\phi}_{\gamma}
\end{split}
\end{equation} 
This is exactly the transformation that we used in the paper. It can be seen that for special positive value of $g^{*}_{z} = \frac{8\pi \sqrt{2}}{3}(\frac{v_0}{3}+\frac{v_1}{4})$, called Toulouse point both of the above conditions are met. Canonical transformation that takes the Hamitonian to the Toulouse point then become:
\begin{equation}
    \begin{pmatrix}
        \phi_{m} \\
        \bar{\phi}_{s}
    \end{pmatrix}=
    \begin{pmatrix}
        \sqrt{2} & -1\\
        -1 & \sqrt{2}
    \end{pmatrix}
    \begin{pmatrix}
        \phi_{\gamma} \\
        \bar{\phi}_{J}
    \end{pmatrix}, \qquad \qquad
    \begin{pmatrix}
        \phi_{s} \\
        \bar{\phi}_{m}
    \end{pmatrix}=
    \begin{pmatrix}
        \sqrt{2} & -1\\
        -1 & \sqrt{2}
    \end{pmatrix}
    \begin{pmatrix}
        \phi_{J} \\
        \bar{\phi}_{\gamma}
    \end{pmatrix}
\end{equation}
And, as discussed in the paper the Hamiltonian obtained in this way, using the above transformation has the necessary symmetry of the IR fixed point.

\subsection{Ground state degeneracy of Abelian groups}
Take $\phi$ to be a chiral left-mover with commutation relation $[\phi(x),\phi(y)]=-i\pi\sgn(x-y)$. Generally we have
\be
e^{is_1\phi(x)}e^{is_2\phi(y)}=e^{is_2\phi(y)}e^{is_1\phi(x)}e^{-s_1s_2[\phi(x),\phi(y)]}=e^{is_2\phi(y)}e^{is_1\phi(x)}e^{is_1s_2\pi \sgn(x-y)}
\ee
So we see that if we chose $s_1s_2=1/2$, they ``half'' anti-commute. 
\subsubsection{SU(2)$_1$ case}
We imagine interaction of the type
\be
{\cal H}_j=\frac{2\pi v}{3}(\vec J_{R,j}^{\hh2}+\vec J_{L,j}^{\hh2})+\lambda (J_{R,j}^+J^-_{L,j+1}+h.c.)
\ee
In this case, we choose 
\be
J^+=e^{i\sqrt{2}\phi}, \andd \xi=e^{\frac{i}{\sqrt 2}\phi}
\ee
$J$ is a current operator and $\xi$ is a primary field of SU(2)$_1$. We can think of $J^+$ creating spin-1 excitations made of two spinons $\xi$, each creating spin-1/2 excitations. For $x\neq y$ they obey the algebra
\be
J^\pm(x)J^-(y)=J^-(y)J^\pm(x), \qquad J^+(x)\xi(y)=e^{i\pi\sgn(x-y)}\xi(y)J^+(x), \qquad \zeta(x)\zeta(y)=\zeta(y)\zeta(x)e^{i\frac{\pi}{2}\sgn(x-y)}.
\ee
Therefore, $\zeta$ is an (abelian) anyon itself and mutually anyonic with $J^\pm$. This is essentially a bosonic (even) $m=2$ FQH state. Again, we can define the string operators
\be
\Gamma_x^{J}=\prod_i \bar J^-_i(x_i)J^+_{i+1}(x_i),\quad \Gamma_y^{J}=J^-_j(0)J^+_j(L),\qquad
\Gamma_x^{\zeta}=\prod_i\bar\zeta\dn_i(x'_i)\zeta\dg_{i+1}(x'_i),\quad \Gamma_y^{\zeta}=\zeta\dn_j(0)\zeta\dg_j(L), 
\ee
We can certainly probe the Aharonov bohm by the $\Gamma_{x,y}$ operators and the commutation relations
We have the usual relations
\be
[\Gamma^J_x,\Gamma^J_y]=0, \andd [\Gamma^J_x,H_{int}]=[\Gamma^J_y,H_{int}]=0, \andd [\Gamma_x^{\zeta},\Gamma_y^J]=[\Gamma_y^{\zeta},\Gamma_x^J]=0
\ee
However, as seen from last two relations, $\Gamma^{\zeta}$ operators do not change $\Gamma^J$ quantum numbers. Rather,
\be
[\Gamma_x^{\zeta},H_{int}]=[\Gamma_y^{\zeta},H_{int}]=0, \andd \{\Gamma_x^{\zeta},\Gamma_y^{\zeta}\}=0
\ee
Therefore, we must take one of the $\Gamma^\zeta_1\ket{\pm}=\pm\ket{\pm}$ as the labeling and the other one $\Gamma^\zeta_2$ as the switching operator. The ground state degeneracy is 2. Note that $\Gamma^J_{x,y}=[\Gamma^\zeta_{x,y}]^2$ is not independent of the $\Gamma^\zeta_{x,y}$.

\subsection{SO(N) Group with $N$ even}

Consider the operators $\chi^a$ with $a=1\dots N=2L$ where $N$ is an even number. These are a set of 2L dimension-1/2 fermions  which transform into each other as a vectorial representation of SO(N), i.e. $\chi^a\to R^a_{\hh b}\chi^b$. We can bosonize this set by introducing $L$ bosons so that $\chi'_r=\chi^{2r-1}=\sqrt{2}\cos\phi_r$ and $\chi''_r=\chi^{2r}=\sqrt{2}\sin\phi_r$. For $N\ge 2$, $\pi_1[SO(N)]=\bb Z_2$ meaning that SO(N) has a double cover, known as Spin(N). For example for SO(5) this is Spin(5)$\sim$ SP(4). This cover group admits a spinorial representation. To see this, we introduce $2^L$ new operators in terms of bosons $\phi_r$:
\be
\omega^{\{\alpha_r\}}=e^{\frac{i}{2}\vec\alpha\cdot\vec\phi}, \qquad \alpha_r=\pm 1
\ee
We denote these operators with $\omega^{\alpha}$, bearing in mind that $\alpha$ is a vector with $\alpha=1\dots 2^N$ possible values. These operators have the dimension $[\omega^{\vec\alpha}]=N/16$ equivalent to the product of $N$ chiral twist operators from the N Ising sectors. For any $r$ we can find two $\alpha$ and $\beta$ such that $\phi_r=\frac{1}{2}(\vec\beta-\vec\alpha)\cdot\vec\phi$. Therefore, $\chi_a$ operators can be expressed in terms of $\omega_\alpha$ operators, so that quite generally we can write
\be
\chi^a=\omega{^\alpha}\dg {\cal T}^a_{\alpha\beta}\omega^\beta, \qquad \label{eq2}
\ee
with $\omega$ operators transforming as the spinorial (fundamental) representation of Spin(N), i.e. $\omega^\alpha\to U^{\alpha}_{\hh\beta}\omega^\beta$ where $U\dg [{\cal T}^a]U=R^a_b{\cal T}^b$. Note that since $\omega{^\alpha}\dg=\omega^{-\alpha}$, Eq.\,\pref{eq2} is invariant under $\omega^\alpha\to-\omega^\alpha$ and therefore, each element of SO(N) corresponds to (only) two elements in Spin(N). 

In CFT these are the spinorial chiral primaries of the SO(N) WZW model. The commutation relation of these two operators can be found from the Baker-Campbell-Hausdorff (BCH) formula
\be
e^{\frac{i}{2}\vec\alpha\cdot\vec\phi(y)}e^{ib\phi_\ell (x)}=e^{ib\phi_\ell (x)}e^{\frac{i}{2}\vec\alpha\cdot\vec\phi(y)}e^{\frac{1}{2}\sum_r\alpha_r b[\phi_\ell(x),\phi_r(y)]}
\ee
which using the commutation relation $[\phi_m(x),\phi_n(y)]=-i\pi\delta_{nm}\sgn(x-y)$ becomes
\be
\omega^\alpha(y)e^{ib\phi_\ell (x)}=-i\sgn(b\alpha_\ell)\sgn(x-y)e^{ib\phi_\ell (x)}\omega^\alpha(y),\qquad b=\pm1
\ee
On the other hand, for the right-movers
\be
\bar\omega^\alpha(y)e^{ib\bar\phi_\ell (x)}=+i\sgn(b\alpha_\ell)\sgn(x-y)e^{ib\bar\phi_\ell (x)}\bar\omega^\alpha(y),\qquad b=\pm1
\ee
It follows from the appearance of $\sgn(b)$ that commutation with $\omega_\alpha$ transforms $\chi_{2r}$ into $\chi_{2r-1}$ and vice versa: 
\be
\matl{\omega^\alpha(y)\chi_{2r-1}(x)=\sgn(\alpha_r)\sgn(x-y)\chi_{2r}(x)\omega^{\alpha}(y)\\
\omega^\alpha(y)\chi_{2r}(x)=-\sgn(\alpha_r)\sgn(x-y)\chi_{2r-1}(x)\omega^{\alpha}(y)}, \qquad
\matl{\bar\omega^\alpha(y)\bar\chi_{2r-1}(x)=-\sgn(\alpha_r)\sgn(x-y)\bar\chi_{2r}(x)\bar\omega^{\alpha}(y)\\
\bar\omega^\alpha(y)\bar\chi_{2r}(x)=\sgn(\alpha_r)\sgn(x-y)\bar\chi_{2r-1}(x)\bar\omega^{\alpha}(y)}\label{eq6}
\ee
However, doing this operation twice cancels out such factors and we find
\be
\omega^\alpha(y)e^{ib\phi_\ell (x_1)}e^{-ib\phi_\ell (x_2)}=\sgn(x_1-y)\sgn(x_2-y)e^{ib\phi_\ell (x_1)}e^{-ib\phi_\ell (x_2)}\omega^\alpha(y)
\ee
or
\be
\omega^\alpha(y)\bar\omega{^\alpha}\dg(y)e^{ib\bar\phi_\ell (x_1)}e^{-ib\phi_\ell (x_2)}=\sgn(x_1-y)\sgn(x_2-y)e^{ib\bar\phi_\ell (x_1)}e^{-ib\phi_\ell (x_2)}\omega^\alpha(y)\bar\omega{^\alpha}\dg(y)
\ee
This form implies that denoting $\vec \chi\cdot\vec\gamma=\sum_a\chi^a\gamma^a$ we have
\be
\omega^\alpha(y)[\chi'_r(x_1)\chi'_r(x_2)+\chi''_r(x_1)\chi''_r(x_2)]\cdot\vec \chi(x_2)=\sgn(x_1-y)\sgn(x_2-y)\vec\chi(x_1)\cdot\vec\chi(x_2)\omega^\alpha(y)
\ee
and
\be
\omega^\alpha(y)\bar\omega{^\alpha}\dg(y)\vec{\bar\chi}(x_1)\cdot\vec\chi(x_2)=\sgn(x_1-y)\sgn(x_2-y)\vec{\bar\chi}(x_1)\cdot\vec\chi(x_2)\omega^\alpha(y)\bar\omega{^\alpha}\dg(y)
\ee
In particular, defining the SO(N) symmetric operators on an arbitrary chiral wire $j$
\be
\Gamma_y^1=i\vec\chi_j(0)\cdot\vec\chi_j(L), \andd \eps_j(x)=i\vec{\bar \chi}_j(x)\cdot\vec \chi_{j+1}(x)
\ee
and a family of tensorial operators
\be
\Gamma_x^{1/2}=\sum_{\{\alpha,\beta\}}{\cal N}_{\{\alpha,\beta\}}\prod_j\bar\omega{^{-\alpha_j}_j}(x_j)\omega^{\beta_j}_{j+1}(x_j)
\ee
it follows that irrespective of the tensor ${\cal N}_{\{\alpha,\beta\}}$,
\be
\Gamma_x^{1/2}\Gamma_y^{1}=-\Gamma_y^{1}\Gamma_x^{1/2}, \qqquad \Gamma_x^{1/2}\eps_j(x)=\eps_j(x)\Gamma_x^{1/2} \so \Gamma_x^{1/2}{\cal H}={\cal H}\Gamma_x^{1/2}
\ee
Also, from Eq.\,\pref{eq6} we see that $\Gamma_x^{1/2}$ introduces a discontinuity $\sgn(x-x_j)$ into each $\chi_j^a(x)$ or $\bar\chi^a_j(x)$, and at the same time shuffling $\chi^{2r}\lr\chi^{2r-1}$. While the precise position of sign change is gauge-dependent and can be moved via a gauge transformation, this changes the boundary condition over these Majorana fields from anti-periodic to periodic or vice verse. Likewise, we define the operators
\be
\Gamma_x^1=\sum_{\{a,b\}}{\cal M}_{\{a,b\}}\prod_ji\bar\chi_j^{a_j}(x_j)\chi_{j+1}^{b_j}(x_j), \qqquad \Gamma_y^{1/2}=\sum_{\alpha}\omega^{-\alpha}	_j(0)\omega^\alpha_{j}(L)
\ee
Particularly for ${\cal M}_{\{a,b\}}=\prod_j\delta_{a_jb_j}$, we have $\Gamma_x^1=\prod_j\eps_j(x_j)$. Using the fact that $\beta=\pm 1$ drops out of the equation
\be
\omega^{\alpha_1}(y_1)\omega^{\alpha_2}(y_2)e^{i\beta\phi_\ell(x)}=-\sgn(\alpha_1^\ell \alpha_2^\ell)\sgn(x-y_1)\sgn(x-y_2)e^{i\beta\phi_\ell(x)}\omega^{\alpha_1}(y_1)\omega^{\alpha_2}(y_2),
\ee
we conclude that irrespective of the tensor ${\cal M}$,
\be
\Gamma_y^{1/2}\chi^a_j(x)=-\chi^a_j(x)\Gamma_y^{1/2}, \orr \Gamma_y^{1/2}\chi^a_j(x)\Gamma_y^{1/2}=-\chi^a_j(x)
\ee
This means that
\be
\Gamma_y^{1/2}{\cal H}\Gamma_y^{1/2}={\cal H}, \andd \Gamma_y^{1/2}\Gamma_x^{1}=-\Gamma_x^{1}\Gamma_y^{1/2}, \qquad \text{but} \qquad
\Gamma_y^{1/2}\Gamma_y^{1}=\Gamma_y^{1}\Gamma_y^{1/2}.
\ee
This algebra is used in the paper to span the degenerate ground state manifold.

\ew

\end{document}